\documentclass[prd,showpacs,twocolumn,preprintnumbers,nofootinbib]{revtex4}

\usepackage{amsmath,amssymb,amsfonts,amsthm}
\usepackage{bm}
\usepackage{graphics}

\newcommand{\be}{\begin{equation}}
\newcommand{\ee}{\end{equation}}
\newcommand{\ba}{\begin{eqnarray}}
\newcommand{\ea}{\end{eqnarray}}
\def\bs{\begin{subequations}}
\def\es{\end{subequations}}
\def\a{\alpha}
\def\b{\beta}
\def\g{\gamma}
\def\e{\epsilon}
\def\om{\omega}

\def\s{\sigma}
\def\cC{{\cal C}}

\def\cH{{\cal H}}
\def\cL{{\cal L}}

\def\cM{{\cal M}}

\def\cK{{\cal K}}
\def\p{\partial}
\newcommand{\Eq}[1]{(\ref{#1})}

\begin{document}

\title{Barbero--Immirzi field in canonical formalism of pure gravity}
\author{Gianluca Calcagni}
\email{gianluca@gravity.psu.edu}
\author{Simone Mercuri}
\email{mercuri@gravity.psu.edu}
\affiliation{Institute for Gravitation and the Cosmos, Department of Physics,\\ The Pennsylvania State University,
104 Davey Lab, University Park, Pennsylvania 16802, USA}
\date{February 5, 2009}
\begin{abstract}
The Barbero--Immirzi (BI) parameter is promoted to a field and a canonical analysis is performed when it is coupled with a Nieh--Yan topological invariant. It is shown that, in the effective theory, the BI field is a canonical pseudoscalar minimally coupled with gravity. This framework is argued to be more natural than the one of the usual Holst action. Potential consequences in relation with inflation and the quantum theory are briefly discussed.
\end{abstract}

\preprint{Phys. Rev. D {\bf 79}, 084004 (2009) \qquad [arXiv:0902.0957]}
\maketitle


\section{Introduction}

Loop quantum gravity (LQG) \cite{rov97,thi01,thi02,AL} aims to quantize the gravitational interaction in a rigorous and consistent way. At classical level, it relies on the Hamiltonian formulation of gravity through Ashtekar--Barbero canonical variables \cite{ash86,ash87,bar95} and features a real constant $\g$ (or $\b\equiv-1/\g$) called the Barbero--Immirzi (BI) parameter \cite{bar95,imm96,RT}. Its value can be constrained by the computation of the entropy of nonrotating black-hole isolated horizons \cite{ABCK,ABK} but otherwise it is arbitrary. Many authors have contributed to clarify the physical origin of the BI parameter \cite{RT,GOP,mer07,DKS} and suggestions have come from studying the interaction of gravity with fermions \cite{FreMinTak05,Ran05,PerRov06,mer06,BojDas08}. 

Recently, it has been proposed to promote $\b$ to a field $\b(x)$ in a Holst-type action and study the resulting dynamics \cite{TY}; the same system was considered also in \cite{TK} and, actually, in an older publication \cite{CDF}. Classically, when the so-called second Cartan structure equation is reinserted into the action and the effective dynamics is extracted, this system turns out to be equivalent to one with a pure gravitational (Einstein--Hilbert) sector and a decoupled scalar field.

The first goal of this paper is to clarify the parity properties of the BI field, which were not recognized previously. This can be done in a straightforward way by decomposing torsion into irreducible components. It is explicitly shown that $\b$ must be a pseudoscalar in order to preserve the transformation properties of these components under the local Lorentz group. 

The Holst action, however, is not the most natural starting point. In fact, $\b$ generates torsion and one should expect the presence of torsion terms already at the action level. As previously argued in \cite{mer06,mer2,mer07}, the Holst framework has to be generalized in order to deal with Riemann--Cartan space-time. This is achieved by completing the Holst term with a torsion part so that the net coupling with $\b$ is nothing but the Nieh--Yan density \cite{NY,CZ,mer06,mer2,mer07,nie07,DKS}. 

In other words, the observation, made by Holst \cite{hol95}, that the Hilbert--Palatini action can be generalized by adding a new term (which, in the case $\beta={\rm const}$, does not affect the classical dynamics of pure gravity) can be extended to torsional space-times \cite{mer06,DKS}. Interestingly enough, in the case $\beta=\beta(x)$ coupled with the Nieh--Yan invariant, we shall see that the BI field becomes a real \emph{canonical} pseudoscalar field. The pseudoscalar nature of the $\b$ field can be demonstrated by noting that the axial component of torsion is proportional to the partial derivative of $\beta$ itself. 

All these results can be obtained in the Lagrangian formalism, but for quantization purposes it is natural to analyze the $\b$ field also within the canonical formalism \cite{dir50} (see \cite{HT} for an introduction). In the standard quantization procedure, Dirac brackets are promoted to commutators and first-class constraints to operators acting on a suitable Hilbert space.\footnote{Other quantization schemes are possible, for instance after solving some of the first-class constraints; this could lead to an altogether different theory.} This motivates us to study the BI field in the first-order Hamiltonian formalism in the absence of matter, in both the Holst and Nieh--Yan case. As expected, the canonical system is fully equivalent to the second-order effective action for a scalar field minimally coupled with Einstein--Hilbert gravity (in agreement with \cite{TY,TK} in the Holst case).

The quantum theory provides us with yet another reason to inspect the Nieh--Yan case. Since $\b$ is now coupled with a topological invariant (while the Holst term vanishes only ``on half shell,'' i.e., when the second Cartan structure equation is solved), one could ask whether it plays a role analog to the $\theta$ parameter in QCD \cite{mer07,DKS} (see also \cite{GOP}; an early proposal for a CP-violating mechanism wherein the Barbero--Immirzi parameter was involved, still as a constant, can be found in \cite{ale05}). In a companion paper \cite{Mer09}, it is argued that the BI parameter must be a field in order to absorb a divergent chiral anomaly \cite{CZ,Soo98} in the presence of fermions. 

Another justification for focussing some attention to the Hamiltonian framework is to clarify in what sense this system (either Holst or Nieh--Yan) is a scalar-tensor theory. In fact, since the $\b$ field is coupled to gravity in the original action, this seems to be a particular example of a Brans--Dicke theory \cite{BD,Dic61}. In these models, one can make a conformal transformation of the metric such that the theory in the ``Jordan frame,'' where the coupling between gravity and the scalar field is nonminimal, is mapped onto one in the ``Einstein frame,'' where the coupling is minimal \cite{mae89}. Physically the two theories are inequivalent, as one can take either frame as the one where distances are measured. Moreover, in the presence of matter the conformal transformation changes the coupling between the scalar and other fields, thus violating the strong equivalence principle. However, on one hand in the effective action of the BI model no trace of the nonminimal coupling is left, and on the other hand it is not obvious how a scalar-tensor coupling is translated into Hamiltonian language. The role of the Nieh--Yan term in the presence of spinor matter fields has been studied in \cite{mer06} and more recently in  \cite{mer07} (see also \cite{DKS}), where the BI parameter has been argued to have a topological origin. Here we do not take spinors into account but claim that, starting from an action containing the Nieh--Yan term and promoting the BI parameter to a field, we obtain a more natural effective action in which $\b$ is canonical.


The paper is organized as follows. Since the Holst case is by far the most widely considered in the literature, we will compare it with the Nieh--Yan case in a step-by-step fashion. The fundamental action with a Holst or a Nieh--Yan term is analyzed in the Lagrangian formalism in Secs.~\ref{lana} and \ref{nylag}, respectively. The Hamiltonian formalism is inspected in Sec.~\ref{hami}, where we quote the main results. The reader unfamiliar with constrained Hamiltonian systems in the presence of torsion can consult the appendix for a pedagogical introduction and a full derivation of the constraints. Section \ref{conc} is devoted to the discussion of the achieved results and future directions. 

In the following, the space-time signature is $({+}{-}{-}{-})$. Repeated upper-lower indices are summed over. We set $8\pi G=1$.


\section{Lagrangian formulation}

Starting from the Lagrangian Holst theory, we calculate the effective action by introducing the irreducible components of torsion and solving their equations of motion. By reintroducing the obtained solutions into the action, we calculate the effective action, demonstrating that the BI field decouples from gravity and also showing its pseudoscalar nature. It is worth stressing that the field $\beta$ is characterized by a complicated kinetic term, which can be recast in the standard way by a simple change of variables, as already noticed in \cite{TY,TYtalk,TK}. Unfortunately, this field redefinition would lead to a quite unnatural coupling between the pseudoscalar and spinor matter.

Then we generalize the Holst action by introducing a new torsion-torsion term in the action. This is motivated by some geometrical arguments suggested by the fact that the Holst modification of the Hilbert--Palatini action is not completely general. In fact, it contains only one of the two terms belonging to a well-known topological density called the Nieh--Yan 4-form. The Nieh--Yan density is linear in the curvature and contains a torsion-torsion term, which can play an important dynamical role in the case torsion does not vanish, i.e., in dynamical systems in which a source for torsion is present. It goes without saying that the new action containing the Nieh--Yan term reduces to the usual Holst action for torsion-free gravity and constant BI parameter. 


\subsection{Holst case}\label{lana}

Let $(\cM^4,g_{\mu\nu})$ be a four-dimensional space-time manifold $\cM^4$ locally equipped with a metric $g_{\mu\nu}$. The tangent space $T_x\cM^4$ is isomorphic to Minkowski space and we can define the one-to-one map $e:\cM^4\rightarrow T_x \cM^4$ which sends tensor fields from the manifold to the Minkowskian tangent space. This map, generally called tetrad or vierbein, is a local reference system for the space-time, physically representing the gravitational field. Its relation with the metric $g_{\mu\nu}$ is summarized in the following formul\ae\:
\begin{equation}\label{tef}
g_{\mu\nu}=\eta_{a b}e^{a}_{\ \mu}e^{b}_{\ \nu}\,,\qquad e^{\ a}_{\mu}e_a^{\ \nu}=\delta^{\nu}_{\mu}\,,\qquad e_a^{\ \mu}e^{\ b}_{\mu}=\delta^{b}_{a}\,,
\end{equation}
where both Greek and Latin indices run from $0$ to $3$ and transform, respectively, under general coordinate and local Lorentz transformations. The tetrad fields incorporate all the metric properties of space-time, but the converse is not true. In fact, due to (manifest) local Lorentz invariance, there are infinitely many realizations of the local basis reproducing the same metric tensor. This is also the reason why there are more components in the tetrads than in the metric field, the difference being exactly six, which is the number of free parameters of the $SO(3,1)$ group representing Lorentz transformations on the Minkowski tangent space.

The action for gravity can be rewritten in terms of the tetrad fields and the Lorentz-valued spin connection $\om^{ab}_{\ \ \mu}$, which will be considered as an independent field according to the Palatini formulation. The usual Hilbert--Palatini action will be generalized to contain the so-called Holst term \cite{hol95} and the Barbero--Immirzi parameter promoted to a field \cite{TY,CDF}:
\ba
S&=&\int\limits_{\cM^4} d t L\nonumber\\
&=&-\frac{1}{2}\int\limits_{\cM^4}d^4x\,^{(4)}\!e\,e_{a}^{\ \mu}e_{b}^{\ \nu}\left(R_{\ \ \mu\nu}^{a b}-\frac{\b}{2}\,\e^{ab}_{\ \ c d}\,R_{\ \ \mu\nu}^{c d}\right)\,,\nonumber\\\label{act}
\ea
where ${}^{(4)}\!e\equiv \det (e^\mu_a)$ is the determinant of the tetrad, $\b=\b(x)$ is the BI field, and $R_{\ \ \mu\nu}^{a b}=2\partial_{[\mu}\om_{\ \ \nu]}^{a b}+2\omega^{a}_{\ c [\mu}\omega_{\ \ \nu]}^{c b}$ is the Riemann curvature associated with $\omega^{a b}_{\ \ \mu}$.\footnote{Square brackets denote antisymmetrized indices, $X_{[\mu\nu]}=\frac{1}{2}\left(X_{\mu\nu}-X_{\nu\mu}\right)$.}

It appears immediately clear that the action (\ref{act}) is not equivalent to the Hilbert--Palatini action, differently from the case $\b= {\rm const}$. The reason is that the second Cartan structure equation is affected by the presence of the BI field; in particular, a torsion contribution depending on the derivative of the BI field will appear in the spin connection. As a consequence, the Bianchi cyclic identity generalizes too, assuming the form
\begin{equation}
R_{\ b [\mu\nu}^{a}e_{\ \rho]}^b=D_{[\mu}T^a_{\ \nu\rho]}\neq 0\,,
\end{equation}
where $D_{\mu}$ is the covariant derivative operator made with the Lorentz-valued spin connection $\om^{ab}$ and $T^{a}_{\ \mu\nu}$ is the \emph{torsion tensor}, which depends on the derivative of $\b$ in this specific case. Then, the Holst term no longer vanishes on half shell, unless $\b$ is a constant. This fact has interesting dynamical consequences. We are going to demonstrate that the BI field, through the torsion tensor, decouples from the gravitational sector of the theory and plays the role of an independent (pseudo)scalar field \cite{CDF}. 

We begin by studying the dynamics described by Eq.~\Eq{act} from a Lagrangian point of view. It is convenient to split the Lorentz spin connection in a torsionless part $\bar{\om}^{ab}$ (Ricci connection, which obeys the homogeneous structure equation) plus the \emph{contortion} 1-form $\cK^{ab}$ \cite{HHKN}:
\be\label{deco}
\om_{\ \ \mu}^{ab}=\bar{\om}_{\ \ \mu}^{ab}+\cK^{ab}_{\ \ \mu}\,,
\ee
where the contortion tensor
\be\label{ccc}
\cK^{ab}_{\ \ \mu}=e_{\ \nu}^a e_{\ \rho}^b \cK^{\nu\rho}_{\ \ \mu}\,,\qquad \cK^{\nu\rho}_{\ \ \mu} =-\cK^{\rho\nu}_{\ \ \mu}\,,
\ee
is related with the torsion $T^{\nu}_{\ \rho\mu}=-T^{\nu}_{\ \mu\rho}$ by
\be\label{conto}
\cK^{\nu}_{\ \rho\mu}=\tfrac12 (T^{\nu}_{\ \rho\mu}-T^{\ \nu}_{\rho\ \mu}-T^{\ \nu}_{\mu\ \rho})\,.
\ee
The action equation \Eq{act} (integration domain omitted) reads
\ba
S &=&-\frac{1}{2}\int d^4x\,^{(4)}\!e\,e_{a}^{\ \mu}e_{b}^{\ \nu}\bar{R}_{\mu\nu}^{\ \ \ a b}\nonumber\\
&&-\frac{1}{2}\int d^4x\,^{(4)}\!e\,e_{a}^{\ \mu}e_{b}^{\ \nu}\left(\cK^{a}_{\ c\mu}\cK^{cb}_{\ \ \nu}-\cK^{a}_{\ c\nu}\cK^{cb}_{\ \ \mu}\right)\nonumber\\
&&+\frac{1}{2}\int d^4x\,^{(4)}\!e\,e_{a}^{\ \mu}e_{b}^{\ \nu}\b\,\e^{ab}_{\ \ cd}\p_{[\mu} \cK^{cd}_{\ \ \nu]}\nonumber\\
&&+\frac{1}{4}\int d^4x\,^{(4)}\!e\,e_{a}^{\ \mu}e_{b}^{\ \nu}\b\,\e^{ab}_{\ \ cd}\nonumber\\
&&\qquad\qquad\times\left(\cK^{c}_{\ f\mu}\cK^{fd}_{\ \ \nu}-\cK^{c}_{\ f\nu}\cK^{fd}_{\ \ \mu}\right),
\ea
where the term $\e^{ab}_{\ \ c d}\,e_a^{\ \mu}e_b^{\ \nu}\bar{R}_{\ \ \mu\nu}^{c d}$ vanishes because of the Bianchi cyclic identity $e_{b[\rho} \bar{R}_{\ \ \mu\nu]}^{ab}=0$ and total divergences have been dropped out. 

It will be particularly convenient to split the torsion into its irreducible components in accordance with the Lorentz group \cite{McC92,BOS,HPS,CLS}:
\be\label{decot}
T_{\mu\nu\rho}=\frac13\,\left(T_\nu g_{\mu\rho}-T_\rho g_{\mu\nu}\right)-\frac16\,\e_{\mu\nu\rho\s}S^\s+q_{\mu\nu\rho}\,,
\ee
where
\be
T_\mu\equiv T^\nu_{\ \mu\nu}
\ee
is the \emph{trace vector},
\be
S_\mu \equiv\e_{\nu\rho\s\mu}T^{\nu\rho\s}
\ee
is the \emph{pseudotrace axial vector}, and the antisymmetric tensor $q_{\mu\nu\rho}$ is such that $q^\nu_{\ \rho\nu}=0=\e^{\mu\nu\rho\s}q_{\mu\nu\rho}$. Equation \Eq{act} can be rewritten as
\ba
S&=&-\frac{1}{2}\int d^4x\,^{(4)}\!e\left[e_{a}^{\ \mu}e_{b}^{\ \nu}\bar{R}_{\mu\nu}^{\ \ \ a b}+\frac{\b}{2}\bar{\nabla}_\mu S^\mu+\frac{1}{24}S_\mu S^\mu\right.\nonumber\\
&&\qquad\qquad\qquad-\frac23 T_\mu T^\mu+\frac{\b}3T_\mu S^\mu\nonumber\\
&&\qquad\qquad\qquad\left.+\frac12q_{\mu\nu\rho}q^{\mu\nu\rho}+\frac{\b}{2}\,\e_{\mu\nu\rho\s}q_{\tau}^{\ \mu\rho}q^{\tau\nu\s}\right]\,,\nonumber\\\label{inters}
\ea
where $\bar{\nabla}_\mu$ is the torsionless and metric-compatible covariant derivative. By varying the action with respect to the irreducible components of torsion $S^{\mu}$, $T^{\nu}$, and $q^{\rho\sigma\tau}$, we obtain respectively:
\ba
\frac12\partial_{\mu}\b-\frac{1}{12}S_{\mu}-\frac{1}{3}\beta T_{\mu}&=&0\,,
\\
\beta S_{\nu}-4T_{\nu}&=&0\,,
\\
q_{\mu\nu\rho}+\b \e_{\nu\sigma\rho\tau}q_{\mu}^{\ \sigma\tau}&=&0\,.
\ea
The solutions of the above system of equations can be easily calculated:
\ba
S_{\mu}&=&\frac{6}{1+\b^2}\partial_{\mu}\b\,,\label{s11}
\\
T_{\nu}&=&\frac{3}{2}\,\frac{\b}{1+\b^2}\partial_{\nu}\b\,,\label{t1}
\\
q_{\mu\nu\rho}&=&0\,.
\ea
By reinserting the solutions into the action (\ref{inters}), we obtain the expected form of the 
effective action:
\ba
S_{\rm eff}&=&-\frac{1}{2}\int d^4x\,^{(4)}\!e\,e_{a}^{\ \mu}e_{b}^{\ \nu}\bar{R}_{\mu\nu}^{\ \ \ a b}\nonumber\\
&&+\frac{3}{4}\int d^4x\,^{(4)}\!e\,\frac{1}{1+\b^2}\,\p_a\b\p^a\b\,.\label{lagr}
\ea
The system is therefore equivalent to Hilbert--Palatini torsion-free gravity plus a massless scalar field with a nonstandard kinetic term. Defining the new field $\phi$ as 
\be\label{defphi}
\phi=\sqrt{3}\sinh^{-1}\b\,,
\ee
the nonstandard term in Eq.~(\ref{lagr}) can be reabsorbed to obtain
\be
S_{\rm eff}=-\frac{1}{2}\int d^4x\,^{(4)}\!e\left(e_{a}^{\ \mu}e_{b}^{\ \nu}\bar{R}_{\mu\nu}^{\ \ \ a b}-\frac{1}{2}\,\p_a\phi\p^a\phi\right)\,.\label{lagr1}
\ee
Now the effective action contains a standard decoupled pseudoscalar field $\phi$, which is connected to the BI field by the relation (\ref{defphi}). For the sake of completeness, it is worth noting that the solution we have obtained passing through the definition of the irreducible components of torsion corresponds to a contortion tensor of the form
\be\label{solK}
\cK^{ab}_{\ \ \mu} = \frac{1}{1+\b^2}\left(\b e^{[a}_{\ \ \mu}\p^{b]}\b-\frac12\, e^{c}_{\ \mu}\e^{ab}_{\ \ cd} \p^d\b\right)\,,
\ee
which agrees with the one obtained in \cite{TY}.


\subsection{Nieh--Yan case}\label{nylag}

As was shown in the previous section, the noncanonical pseudoscalar $\b$ induces contortion in the spin connection. On general grounds one would expect to meet torsion terms already at the level of the fundamental action. Equation \Eq{act} can be generalized to
\ba
S&=&-\frac{1}{2}\int d^4x\,^{(4)}\!e\,e_{a}^{\ \mu}e_{b}^{\ \nu}R_{\mu\nu}^{\ \ \ a b}\nonumber\\
&&-\frac14\int d^4x\,^{(4)}\!e\,\b\left(\e^{\mu\nu\rho\s}\eta_{ab}T^a_{\ \mu\nu} T^b_{\ \rho\s}\right.\nonumber\\
&&\left.\qquad\qquad\qquad-e_{a}^{\ \mu}e_{b}^{\ \nu}\e^{ab}_{\ \ cd}\,R_{\mu\nu}^{\ \ \ cd}\right)\,;\label{actny}
\ea
when $\b$ is constant, the second integral becomes the Nieh--Yan topological invariant, which reduces to a total divergence not affecting the equations of motion. In the presence of torsion it is the natural extension of the Holst term, which is not topological by itself.

Using the same procedure of the previous section we can rewrite the above action as
\ba
S&=&-\frac{1}{2}\int d^4x\,^{(4)}\!e\left[e_{a}^{\ \mu}e_{b}^{\ \nu}\bar{R}_{\mu\nu}^{\ \ \ a b}+\frac{\b}{2}\bar{\nabla}_\mu S^\mu\right.\nonumber\\
&&\left.\qquad+\frac{1}{24}S_\mu S^\mu-\frac23 T_\mu T^\mu+\frac12q_{\mu\nu\rho}q^{\mu\nu\rho}\right] \,.\label{naction}
\ea
The terms $(\b/3)T_\mu S^\mu$ and $(\b/2)\e_{\mu\nu\rho\s}q_{\tau}^{\ \mu\rho}q^{\tau\nu\s}$ featured in Eq.~\Eq{inters} have been cancelled out. By varying the action with respect to the irreducible components of torsion $S^{\mu}$, $T^{\nu}$, and $q^{\rho\sigma\tau}$, we obtain, respectively,
\ba
\partial_{\mu}\b-\frac{1}{6}S_{\mu}&=&0\,,
\\
T_{\nu}&=&0\,,
\\
q_{\mu\nu\rho}&=&0\,.
\ea
After reinserting the solutions above into the action (\ref{naction}), we get the effective action
\ba
S_{\rm eff}&=&-\frac{1}{2}\int d^4x\,^{(4)}\!e\,e_{a}^{\ \mu}e_{b}^{\ \nu}\bar{R}_{\mu\nu}^{\ \ \ a b}\nonumber\\
&&+\frac{3}{4}\int d^4x\,^{(4)}\!e\,\p_a\b\p^a\b\,.\label{lagr2}
\ea
The system is therefore equivalent to Hilbert--Palatini torsion-free gravity plus a massless scalar field. Contrary to the Holst case, $\b$ itself is canonical and there is no need to make a field redefinition.


\section{Hamiltonian formulation}\label{hami}

The same results of the previous section can be obtained in the Hamiltonian framework. Following the Dirac procedure, we calculate the first- and second-class constraints of the Holst theory. The latter can be easily solved, so that the system turns out to be characterized by a set of first-class constraints which reflect rotational and diffeomorphism gauge freedom. The counting of the degrees of freedom shows the presence of a free pseudoscalar field decoupled from gravity. 

In order not to distract the reader with a lengthy derivation of the constraints, we refer to the appendix for notation and details. The phase space is equipped with the symplectic structure
\bs\ba
&&\left\{K_{\a}^{i}(t,{\bf x}),E^{\gamma}_j(t,{\bf x}^{\prime})\right\}=\delta^{\g}_{\a}\delta^{i}_{j}\delta({\bf x},{\bf x}^{\prime})\,,\\
&&\left\{\b(t,{\bf x}),\Pi(t,{\bf x}^{\prime})\right\}=\delta({\bf x},{\bf x}^{\prime})\,,
\ea\es
where all indices are spatial (Greek ones over manifold spatial coordinates), $K_{\a}^{i}= \om^{0i}_{\ \ \a}$ is the extrinsic curvature, $E_i^\a= -e e_i^\a$ is the triad, and $\Pi=e(n\cdot S)/4$ is the momentum of the BI field.
The total Hamiltonian is
\be
H_D=\int d^3x\left(\Lambda^i\mathcal{R}_i+N^\a\cH_\a+N\cH\right),
\ee
where $\Lambda^i$, $N^\a$, and $N$ are Lagrange multipliers. In the Holst case, the rotation, supermomentum, and super-Hamiltonian (first-class) constraints are, respectively,
\bs\ba
\mathcal{R}_i&\equiv & \e_{i j}^{\ \ k}K_{\a}^{j}E^\a_{k}\approx 0\,,\label{rotaz}\\
\cH_\a &\equiv &  2E^{\gamma}_i D_{[\a}K^{i}_{\g]}+\Pi\p_\a\b\approx 0\,,\label{sm}\\
\cH    &\equiv & -\frac{1}{2e}E_i^{\a}E_j^\g\left(\e^{ij}_{\ \ k}R_{\a\g}^k+2K_{[\a}^{i}K_{\g]}^{j}\right)\nonumber\\
&&+\frac{1+\b^2}{3e}\Pi^2-\frac34\frac{e}{1+\b^2}\p_\a\b\p^\a\b\approx 0\,,\label{SH}
\ea\es
where $D_{\a}$ is the covariant derivative in terms of the $SO(3)$-valued spin connection $\Gamma_\a^i=\e^i_{\ j k}\om_\a^{\ j k}/2$ and $R_{\a\g}^k$ is the curvature of $\Gamma$.

In the Nieh--Yan case, the super-Hamiltonian \Eq{SH} is replaced by the simpler
\ba
\cH    &=& -\frac{1}{2e}E_i^{\a}E_j^\g\left(\e^{ij}_{\ \ k}R_{\a\g}^k+2K_{[\a}^{i}K_{\g]}^{j}\right)\nonumber\\
&&+\frac{1}{3e}\Pi^2-\frac34e\,\p_\a\b\p^\a\b\,.\label{SH2}
\ea
The net effect of the Nieh--Yan term is to absorb factors $(1+\b^2)$ in the contributions of the pseudoscalar field, which is now canonical. Notice that the system possesses a shift symmetry $\b\to \b+\b_0$ which is absent in the Holst case.

\section{Discussion}\label{conc}

The technical results of this paper may open up some interesting lines of investigation. 

In either the Holst or the Nieh--Yan case, the canonical approach clearly shows [see Eqs.~\Eq{sm}, \Eq{SH}, and \Eq{SH2}] that the matter-free system under consideration is not equivalent to a scalar-tensor theory in a nontrivial sense. Although one can almost always perform a conformal transformation of a minimally coupled scalar-tensor system to get a Brans--Dicke type theory, in the absence of extra matter this step may be physically justified only in the other direction, i.e., from a scalar-tensor theory to a minimally coupled one. When matter is included, the change of frame would determine different couplings between the matter sector and the scalar field. This happens to be the case, for instance, when fermions are included, but in the Holst case the resulting coupling is rather unnatural \cite{TK}. On the other hand, the scalar-fermion coupling is drastically simplified in the Nieh--Yan case \cite{mer06,MeT}. The naturalness of the action in both the scalar and fermionic sectors indeed makes the Nieh--Yan case more appealing. 

In fact, if $\b$ and the inflaton were identified, then inflation would be reinterpreted as a phenomenon stemming from a breaking of the topological sector of the theory (in \cite{AC1,AC2} a similar claim was made, although from a different physical perspective). However, from the theory we do not have any input as far as a potential for $\b$ is concerned. An insertion by hand would not explain inflation more than any other phenomenological model. If one required CP symmetry to hold, the potential would be restricted to even functions of $\b$; otherwise some CP-violating effect might make its appearance during the evolution of the universe. A natural potential with nontrivial minima might be achieved via a Peccei--Quinn mechanism \cite{Mer09}. At that point one could also ask oneself whether the value of $\b$ found from black hole entropy calculations \cite{ABCK,ABK} is related to a particular vacuum.

Finally, since loop quantum gravity makes extensive use of the Ashtekar--Barbero connection and its conjugate momentum, which allow for a well-defined quantization, it is of interest to reexpress the constraints from $(K,E)$ variables to the latter. It is immediately clear that the most naive generalization of the Ashtekar--Barbero connection to a varying BI parameter, $\tilde A_\a^i \equiv -(1/\b)K_\a^i+\Gamma_\a^i$, would not lead to a canonical algebra \cite{menmar}. For instance, $\left\{A^{i}_{\a},\Pi\right\}\neq 0$, due to the mixing of matter and gravitational degrees of freedom in $A$. There is another way to state this result. The rotation constraint and the saturated compatibility condition combine into the Gauss constraint ${\cal D}_\a E^\a_i\equiv \p_\a E^\a_i+\e_{ij}^{\ \ k}A_\a^j E^\a_k\approx 0$. Taking the Poisson bracket of the Gauss constraint with itself, one can see that the algebra of gauge rotations does not close. In fact, the above definition would break the shift symmetry in the $\b$ field, thus leading to a different theory. Obviously, with the usual definition of the connection with constant $\b_0$,
\be
A_\a^i \equiv -\frac{1}{\b_0}K_\a^i+\Gamma_\a^i\,,\label{A}
\ee
the symplectic structure remains canonical, and one gets the well-known constraint equations with the addition of the scalar sector. Now, one should justify the definition \Eq{A} and explain the relation between the constant $\b_0$ and the BI field. An explicit parametrization of $\b(t,{\bf x})$ in terms of space-time coordinates could shed some light on this issue.



\begin{acknowledgments}
This research was supported in part by NSF Grant No.~PHY0854743, the George A. and Margaret M. Downsbrough Endowment and the Eberly research funds of Penn State. We thank A. Ashtekar for useful comments.
\end{acknowledgments}


\appendix*

\section{Hamiltonian analysis with torsion}


\subsection{Holst case} \label{hol}

Let us consider the action (\ref{inters}) and assume that the space-time $(\cM^4,g_{\mu\nu})$ is globally hyperbolic \cite{HE}. Then, according to Geroch theorem \cite{ger70} (see also \cite{BS}), a global time function $t$ can be chosen in such a way that each surface of constant $t$ is a Cauchy surface and space-time topology is $\cM^4=\mathbb{R}\times\Sigma^3$, where $\Sigma^3$ is any Cauchy surface. On each surface, the metric \Eq{tef} induces a Riemannian metric $h_{\mu\nu}$ defined by the first fundamental form, i.e.
\be
h_{\mu\nu}=g_{\mu\nu}-n_{\mu}n_{\nu}\,,
\ee
where $n^{\mu}$ is the normal vector to $\Sigma^3$. Let $t^{\mu}=t^{\mu}(y)$ be the \emph{time flow} vector field on $\cM^4\ni y$ satisfying $t^\mu\nabla_\mu t=t^\mu\p_\mu t=1$.\footnote{We remark that neither $t$ nor $t^{\mu}$ can be interpreted in terms of physical measurements of time, since one does not know the metric, which is, in fact, the unknown dynamical field in the Einstein theory of gravitation.} The time flow vector field generates a one-parameter group of diffeomorphisms, known as embedding diffeomorphisms, $\phi_t:\mathbb{R}\times\Sigma^3\rightarrow \cM^{4}$, defined as $y(t, {\bf x})\equiv y_t({\bf x})$. This allows to represent space-time as a smooth deformation of the three-dimensional Cauchy surfaces $\Sigma^{3}$ into a one-parameter family of three-dimensional Cauchy surfaces $\Sigma^3_t$. These are described by the parametric equations $y^{\mu}_t=y^{\mu}_t({\bf x})$, where $t$ denotes the hypersurface at different ``times''. A general parametrization can be obtained by introducing the normal and tangential components of the vector field $t^{\mu}(y)$ with respect to $\Sigma^3$. Namely, we define
\be
N\equiv g_{\mu\nu}t^{\mu}n^{\nu}\,,\qquad N^{\mu}\equiv h^{\mu}_{\ \nu}t^{\nu}\,,
\ee
respectively called the \emph{lapse function} and \emph{shift vector}. As a consequence we have
\ba
t^{\mu}(t, {\bf x})&=& \left.\frac{\partial y^{\mu}(t, {\bf x})}{\partial t}\right|_{y(t, {\bf x}) =y_t({\bf x})}\nonumber\\
&=&N(t, {\bf x})\,n^{\mu}(t, {\bf x})+N^{\mu}(t, {\bf x})\,.\label{def}
\ea
By acting with a Wigner boost on the local basis, we can rotate it so that its zeroth component results to be parallel, in each point of $\Sigma^3$, to the normal vector $n_{\mu}$, i.e. $n_\mu=e_{\ \mu}^0$ (implying that the local boost parameter $e_{\ t}^i$ vanishes at each point of $\Sigma^3$). The requirement that this particular choice of the orientation of the local basis be preserved along the evolution fixes the so-called Schwinger or time gauge, the net result being that the action will no longer depend on the boost parameters; also, the local symmetry group is reduced from the initial $SO(3,1)$ to $SO(3)$, which encodes the spatial rotational symmetry. It can be demonstrated that fixing the time gauge into the action does not affect the consistency of the canonical analysis, this procedure being equivalent to a canonical gauge fixing. The action \Eq{inters} can be finally written as follows:
\ba
S&=&-\frac{1}{2}\int d t d^3x\,{}^{(4)}\!e\,\left(\vphantom{\frac12}2e_0^{\ \mu}e_i^{\ \nu}R_{\mu\nu}^{\ \ \ 0i}+e_i^{\ \mu}e_j^{\ \nu}R_{\mu\nu}^{\ \ \ i j}\right.\nonumber\\
&&\left.-\frac12 S^\mu\p_\mu\b+\frac{1}{24}S_\mu S^\mu-\frac23 T_\mu T^\mu+\frac{1}3\b T_\mu S^\mu\right)\nonumber
\\
&=&-\frac{1}{2}\int d t d^3x\,{}^{(4)}\!e\,\left[2\frac{t^\mu-N^\mu}{N}\,e_i^{\ \nu}R_{\mu\nu}^{\ \ \ 0i}\right.\nonumber\\
&&+e_i^{\ \a}e_j^{\ \g}R_{\a\g}^{\ \ \ i j}+(n^\mu n^\nu+h^{\mu\nu})\nonumber\\
&&\left.\times\left(-\frac12 S_\mu\p_\nu\b+\frac{1}{24}S_\mu S_\nu-\frac23 T_\mu T_\nu+\frac{1}3\b T_\mu S_\nu\right)\right]\nonumber
\\
&=&-\int d t d^3x\,e\left\{\vphantom{\frac12}e_i^{\a}\left[t^{\mu}\partial_{\mu}K_\a^i+\om^{0i}_{\ \ \mu}\p_{\a} t^{\mu}-\p_\a\left(t\cdot\om^{i}\right)\right.\right.\nonumber\\
&&\left.-\om_{\ k\a}^{i}(t\cdot\om^{k})
+(t\cdot\om_{\ k}^{i})K_{\a}^{k}\right]\nonumber\\
&&-N^{\alpha} e^{\gamma}_i {}^{(3)}\!R_{\a\g}^{\ \ \ 0 i}+\frac{N}{2}e_i^{\a}e_j^{\g}\left(^{(3)}\!R_{\a\g}^{\ \ \ i j}-2K_{[\a}^{i}K_{\g]}^{j}\right)\nonumber\\
&& -\frac14 (n\cdot S) t^\nu\p_\nu\b +\frac14 (n\cdot S) N^\a\p_\a\b-\frac{N}{4} S^\a \p_\a\b\nonumber\\
&&+N\left[\frac{1}{48}(n\cdot S)^2-\frac13 (n\cdot T)^2+\frac{1}6\b(n\cdot T)(n\cdot S)\right.\nonumber\\
&&\left.\left.+\frac{1}{48}S_\a S^\a-\frac13 T_\a T^\a+\frac{1}6\b T_\a S^\a\right]\right\}\,,\label{act2}
\ea
where we have omitted the bars for torsionless geometrical objects (they will be reinstated only in Sec.~\ref{nylag}) and we set the tensor $q_{\mu\nu\rho}$ to zero since it is nondynamical and does not contribute to the torsion tensor, as was clear from the Lagrangian analysis.\footnote{Setting $q_{\mu\nu\rho}=0$ does not affect the generality of the formulation and it has the advantage of simplifying the canonical formulation, which is rather involved for $3-$tensors like $q_{\mu\nu\rho}$.} The following notation have been used:
\be
K_{\a}^{i}\equiv \om^{0i}_{\ \ \a},
\ee
$t\cdot\om^{i}=t^{\mu}\om^{0i}_{\ \ \mu}$, $t\cdot\om^{jk}= t^{\mu}\om^{jk}_{\ \ \mu}$, $n\cdot S=n^\mu S_\mu$, $n\cdot T=n^\mu T_\mu$, and ${}^{(4)}\!e=Ne$, $e=\det{e^{i}_{\a}}$ being the determinant of the triad. Greek indices $\a,\g,\dots$ from the beginning of the alphabet and Latin indices $i,j,\dots$ from the middle of the alphabet run from $1$ to $3$ and denote, respectively, components transforming under spatial diffeomorphisms and local spatial rotations. The three-dimensional Levi-Civita symbol is defined as $\e_{i j k}\equiv \e_{0ijk}$ and we will often make use of the relation $\e_{ijk}\e^{iln}=\delta_j^n\delta_k^l-\delta_j^l\delta_k^n$.

Now, remembering the definition of the Lie derivative operator on a vector, $\cL_t V_\mu=t^\nu\p_\nu V_\mu+V_\nu\p_\mu t^\nu$, we can rewrite the above action as
\ba
S&=&-\int d t d^3x\,e\left\{\vphantom{\frac12}e_i^{\a}\left[\cL_{t}K_\a^{i}-D_\a\left(t\cdot\om^{i}\right)\right.\right.\nonumber\\
&&\left.+(t\cdot\om_{\ k}^{i})K_{\a}^{k}\right]-2N^{\alpha} e^{\gamma}_i D_{[\a}K^{i}_{\g]}\nonumber\\
&&+\frac{N}{2}e_i^{\a}e_j^{\g}\left(^{(3)}\!R_{\a\g}^{\ \ \ i j}-2K_{[\a}^{i}K_{\g]}^{j}\right)\nonumber\\
&&-\frac14 (n\cdot S) \cL_t\b +\frac14 (n\cdot S) N^\a\p_\a\b-\frac{N}{4} S^\a \p_\a\b\nonumber\\
&&+N\left[\frac{1}{48}(n\cdot S)^2-\frac13 (n\cdot T)^2+\frac{1}6\b(n\cdot T)(n\cdot S)\right.\nonumber\\
&&\left.\left.+\frac{1}{48}S_\a S^\a-\frac13 T_\a T^\a+\frac{1}6\b T_\a S^\a\right]\right\}\,,\label{acts}
\ea
where we introduced the $SO(3)$-valued covariant derivative $D_{\a}(=\bar{D}_\a)$, which can be written as $D_\a V^i=\p_\a V^i+\e^i_{\ jk}\Gamma_\a^j V^k$  on a gauge vector, where
\be
\Gamma_\a^i\equiv\frac12\,\e^i_{\ j k}\om_\a^{\ j k}\,.
\ee
The curvature of $\Gamma$ is defined as
\be\label{curvR}
R^i_{\a\g}\equiv 2\p_{[\a}\Gamma^{i}_{\g]}+\epsilon^i_{\ j k}\Gamma^j_\a \Gamma^k_\g=\frac12\, \e^i_{\ jk}{}^{(3)}\!R_{\a\g}^{\ \ \ jk} \,.
\ee
The next step is the definition of the momenta conjugated to the fundamental variables. Since the Lagrangian is singular, we expect a set of primary constraints to appear. In particular, the only nonvanishing momenta are those conjugated to $K_\a^{i}$ and $\b$:
\bs
\ba
K_\a^{i}\,&:&\qquad\qquad E_i^\a \equiv\frac{\delta S}{\delta \cL_t K_\a^{i}}= -e e_i^\a\,,\label{Eia}\\
\b\,&:&\qquad\qquad \Pi\equiv\frac{\delta S}{\delta \cL_t \b}=\frac14\,e(n\cdot S)\,.
\ea
\es
All the others vanish identically, i.e.
\bs\label{primc}
\ba
e^{\a}_i\,&:&\qquad\qquad\mathcal{P}^i_\a=0\,,\\
\Gamma_\a^{i}\,&:&\qquad\qquad\Pi^\a_{i}=0\,,\\
t\cdot\om^{i}\,&:&\qquad\qquad  \Pi_i= 0\,,\\
t\cdot\om^{ij}\,&:&\qquad\qquad  \Pi_{ij}= 0\,,\\
n\cdot S\,&:&\qquad\qquad  \Pi^{(S)}= 0\,,\\
n\cdot T\,&:&\qquad\qquad  \Pi^{(T)}= 0\,,\\
S^\a\,&:&\qquad\qquad \Pi^{(S)}_\a=0\,,\\
T^\a\,&:&\qquad\qquad \Pi^{(T)}_\a=0\,,\\
N^\a\,&:&\qquad\qquad  \Pi^{(N)}_\a=0\,,\\
N\,&:&\qquad\qquad  \Pi^{(N)}=0\,.
\ea
\es
Here we have not made use of the fact that $\bar\om_{\ \ \a}^{ij}$ is the Ricci spin connection, which depends on the triad field; the canonical analysis will eventually show that it is not an independent variable.

As one can immediately notice, in none of the above conjugated momenta there is the temporal Lie derivative of any of the fundamental variables; so, in principle, all of them should be considered as primary constraints. Therefore the following set of primary constraints has to be imposed:
\bs\label{primary}
\ba
{}^{(K)}\!C^{\a}_i&\equiv& E_i^\a +e e_i^\a\approx 0\,,\label{p1}
\\
\cC&\equiv& \Pi-\frac14\,e(n\cdot S)\approx 0\,,\label{p2}
\\
{}^{(e)}\!C^i_\a&\equiv& \mathcal{P}^i_\a\approx 0\,,\label{p3}
\\
{}^{(\Gamma)}\!C^\a_{i}&\equiv& \Pi^\a_{i}\approx 0\,,\label{po}
\\
C_{i}&\equiv& \Pi_i\approx 0\,,\label{p5}
\\
C_{i j}&\equiv& \Pi_{i j}\approx 0\,,\label{p6}
\\
C^{(b)}&\equiv& \Pi^{(b)}\approx 0\,,\qquad b=S,T,N\,,\label{pstn}
\\
C_{\a}^{(b)}&\equiv& \Pi_\a^{(b)}\approx 0\,,\qquad b=S,T,N\,.\label{pistn}
\ea
\es
The phase space has been equipped with the symplectic structure
\bs\label{commutation relations}
\begin{align}
\left\{K_{\a}^{i}(t,{\bf x}),E^{\gamma}_j(t,{\bf x}^{\prime})\right\}&=\delta^{\g}_{\a}\delta^{i}_{j}\delta({\bf x},{\bf x}^{\prime})\,,
\\
\left\{\b(t,{\bf x}),\Pi(t,{\bf x}^{\prime})\right\}&=\delta({\bf x},{\bf x}^{\prime})\,,
\\
\left\{e^{i}_{\a}(t,{\bf x}),\mathcal{P}^{\g}_j(t,{\bf x}^{\prime})\right\}&=\delta^{\g}_{\alpha}\delta^{i}_{j}\delta({\bf x},{\bf x}^{\prime})\,,
\\
\left\{\Gamma_{\a}^{i}(t,{\bf x}),\Pi^{\gamma}_{j}(t,{\bf x}^{\prime})\right\}&=\delta^{\g}_{\alpha}\delta^{i}_{j}\delta({\bf x},{\bf x}^{\prime})\,,
\\
\left\{t\cdot\om^{i}(t,{\bf x}),\Pi_{k}(t,{\bf x}^{\prime})\right\}&=\delta^{i}_k\delta({\bf x},{\bf x}^{\prime})\,,
\\
\left\{t\cdot\om^{ij}(t,{\bf x}),\Pi_{k l}(t,{\bf x}^{\prime})\right\}&=\delta^{i}_{[k}\delta^{j}_{ l]}\delta({\bf x},{\bf x}^{\prime})\,,
\\
\left\{S(t,{\bf x}),\Pi^{(S)}(t,{\bf x}^{\prime})\right\}&=\delta({\bf x},{\bf x}^{\prime})\,,
\\
\left\{T(t,{\bf x}),\Pi^{(T)}(t,{\bf x}^{\prime})\right\}&=\delta({\bf x},{\bf x}^{\prime})\,,
\\
\left\{N(t,{\bf x}),\Pi^{(N)}(t,{\bf x}^{\prime})\right\}&=\delta({\bf x},{\bf x}^{\prime})\,,
\\
\left\{S^{\alpha}(t,{\bf x}),\Pi_{\g}^{(S)}(t,{\bf x}^{\prime})\right\}&=\delta^{\alpha}_{\g}\delta({\bf x},{\bf x}^{\prime})\,,
\\
\left\{T^{\alpha}(t,{\bf x}),\Pi_{\g}^{(T)}(t,{\bf x}^{\prime})\right\}&=\delta^{\alpha}_{\g}\delta({\bf x},{\bf x}^{\prime})\,,
\\
\left\{N^{\alpha}(t,{\bf x}),\Pi_{\g}^{(N)}(t,{\bf x}^{\prime})\right\}&=\delta^{\alpha}_{\g}\delta({\bf x},{\bf x}^{\prime})\,,
\end{align}
\es
where $\{\cdot,\cdot\}$ are Poisson brackets. Having calculated the conjugated momenta, we can now perform the Legendre dual transformation and extract the canonical Hamiltonian. Since the latter is not uniquely determined because of the primary constraints, we write the Dirac Hamiltonian:
\ba
H_D&=&\int d^3x\left(E_i^\a\cL_t K_\a^{i}+\Pi^{(\b)}\cL_t\b+\lambda^mC_m\right)-L\nonumber\\
&=&\int d^3x\,\left\{\vphantom{\sum_S}E^{\a}_i\left[D_\a\left(t\cdot\om^{i}\right)
-(t\cdot\om_{\ k}^{i})K_{\a}^{k}\right]\right.\nonumber\\
&&+N^\a\cH_\a+N\cH+{}^{(K)}\!\lambda_{\a}^i{}^{(K)}\!C^{\a}_i+\lambda\cC\nonumber\\
&&+{}^{(e)}\!\lambda^{\a}_i\, {}^{(e)}\!C_{\a}^i+{}^{(\Gamma)}\!\lambda_\a^{i} {}^{(\Gamma)}\!C^\a_{i}+\lambda^i C_i+\lambda^{i j}C_{i j}\nonumber\\
&&\left.+\sum_{b=S,T,N}\left[\lambda^{(b)} C^{(b)}+\lambda^\a_{(b)} C_\a^{(b)}\right]\right\},
\ea
where 
\be
\cH_\a \equiv  2E^{\gamma}_i D_{[\a}K^{i}_{\g]}+\Pi\p_\a\b
\ee
is the \emph{supermomentum} and
\ba
\cH &\equiv& -\frac{1}{2e}E_i^{\a}E_j^\g\left(\e^{ij}_{\ \ k}R_{\a\g}^k+2K_{[\a}^{i}K_{\g]}^{j}\right)\nonumber\\
&&+\frac{1}{3e}\Pi^2+\frac23(n\cdot T)\b\Pi\nonumber\\
&&+
e\left[\frac{1}{48}S_\a S^\a-\frac13 (n\cdot T)^2-\frac13 T_\a T^\a\right.\nonumber\\
&&\qquad\left.+\frac{1}6\b T_\a S^\a-\frac{1}{4} S^\a \p_\a\b
\right]\label{sh}
\ea
is the \emph{super-Hamiltonian}, while $\Lambda_m$ and $\lambda_m$ are arbitrary functions. 

As a consistency requirement, the Dirac canonical procedure imposes to calculate the Poisson brackets between the primary constraints and the Dirac Hamiltonian. If they do not vanish on the primary surface for some value of the Lagrange multipliers $\lambda_m$, they must be constrained to vanish. This way, secondary constraints are generated which determine the secondary constraint surface on the phase space \cite{HT}. 

The Poisson brackets between the Dirac Hamiltonian and the first three primary constraints Eqs.~\Eq{p1}, \Eq{p2}, and \Eq{p3} do not generate any secondary constraints: In fact, they can be set to zero by suitably choosing the Lagrange multipliers ${}^{(K)}\!\lambda^\a_i$ and $\lambda^{(S)}$.

For the other primary constraints one gets
\ba
\{C_i,H_D\}     &=& D_\a E_i^\a\,,\label{ga}\\
\{C^{(S)},H_D\} &=& 0\,,\label{s3}\\
\{{}^{(\Gamma)}\!C_i^\a,H_D\}  &=& \e_{ijk}E^{\a j}\left[\vphantom{\frac12}(t\cdot\om^k)-N^{\g}K_\g^k\right.\nonumber\\
&&\left.\qquad\qquad\qquad+E^{\g k}\p_\g\left(\frac{N}e\right)\right]\nonumber\\
&&+N^{\a}\e_{ijk} E^{\g j} K_\g^k+\frac{N}{e}\,\e_i^{\ jk}D_\g\left(E^\a_j E^\g_k\right)\,,\nonumber\\\label{ric1}\\
\{C_{\a}^{(S)},H_D\}  &=& \frac{Ne}2\left(\frac12\p_\a\b-\frac{1}{12}S_\a-\frac13\b T_\a\right)\,,\label{s1}\\
\{C_{\a}^{(T)},H_D\}  &=& \frac{Ne}3\left(2T_\a-\frac12\b S_\a\right)\,,\label{s2}\\
\{C^{(T)},H_D\} &=& \frac23N\left[e(n\cdot T)-\b\Pi\right]\,,\label{s4}\\
\{C_{ij},H_D\}  &=& K_{\a [j}E^\a_{i]}\,,\label{s5}\\
\{C_\a^{(N)},H_D\}    &=& -\cH_\a\,,\label{s6}\\
\{C^{(N)},H_D\} &=& -\cH\,.\label{s7}
\ea
Since we have extracted the torsion components from the full spin connection, the triad obeys the \emph{homogeneous structure equation} $D_{[\a}e^i_{\g]}=0$. Hence, $\Gamma_{\a}^{i}$ is the spatial torsion-free $SO(3)$ spin connection $2\Gamma_{a}^{i}[E]=\e^i_{\ jk}E^{\g j}\nabla_\a E_{\g}^k$, where the covariant derivative $\nabla_\a$ contains the Christoffel symbols, in turn expressed as functions of $E^{\a}_i$ (see \cite{thi01} for the explicit expression). In particular, this implies that the Lagrange multiplier ${}^{(\Gamma)}\!\lambda_\a^{i}$ is determined by the equation of motion of $\Gamma_{a}^{i}[E]$.

On the other hand, the dynamical equations of the canonical variables $S^\a$, $T^\a$, $n\cdot T$, $t\cdot\om^{i}$, $t\cdot\om^{ij}$, $N$, and $N^{\a}$ are completely arbitrary, since each of their Poisson brackets with the full Dirac Hamiltonian is proportional to the associated Lagrange multiplier $\lambda_m$ (the same is true for the equations of motion of $e^\a_i$, $n\cdot S$, and $\Gamma_\a^i$, but as argued above their Lagrange multipliers are no longer arbitrary). Therefore, at this point a useful simplification of this canonical system of constraints can be naturally provided and consists in treating the above subset of canonical variables directly as Lagrange multipliers. This could have been done at the very beginning by inferring which are the dynamical variables and which are the Lagrange multipliers. However, the Dirac procedure does not give us any hint about this classification \emph{ab initio}, so here we have preferred to follow the general procedure and arrive at this conclusion after having calculated the set of primary and secondary constraints.

Equation \Eq{ga} is solved because of the compatibility equation, and $D_\a E_i^\a=0$ strongly (i.e., on all phase space). Equation \Eq{s3}, on the other hand, is a consequence of the fact that $n\cdot S$ disappears from the Dirac Hamiltonian, so that its momentum is preserved by the Hamiltonian flow. Then we can set $\Pi^{(S)}=0$ strongly, as it vanishes initially.

Equations \Eq{ric1}--\Eq{s7} do not contain Lagrange multipliers $\lambda_m$ and do not vanish on the primary surface. Hence they have to be considered as secondary constraints.

Equation \Eq{ric1} has been arranged to isolate three terms. The first includes some of the new Lagrange multipliers and can be made to vanish by definition. The second is proportional to Eq.~\Eq{s5}, so it vanishes weakly. The third term is strongly equal to zero as it is nothing but the compatibility equation. Overall, Eq.~\Eq{ric1} is redundant with other constraints and it will be ignored from now on.

The expression of the new Lagrange multipliers $n\cdot T$, $S_\a$, and $T_\a$ can be easily calculated in order for the secondary constraints Eqs.~\Eq{s1}--\Eq{s4} to vanish:
\ba
n\cdot T &=& \frac1e\,\b\Pi\,,\label{ntsol}\\
S_\a     &=& \frac{6}{1+\b^2}\p_\a\b\,,\label{pseu}\\
T_\a     &=& \frac32\frac{\b}{1+\b^2}\p_\a\b\,.\label{tasol}
\ea
According to Eq.~\Eq{pseu}, $\b$ must be a pseudoscalar. Plugging these expressions into Eq.~\Eq{sh}, we get the reduced set of first-class constraints given by Eqs.~\Eq{rotaz}--\Eq{SH}.

To summarize, the initial complicated system of constraints has been reduced to a set of seven first-class constraints, Eqs.~\Eq{rotaz}--\Eq{SH}, which reflect the gauge freedom of the physical system, i.e., rotation of the local spatial basis and diffeomorphisms of space-time. As regards the canonical variables, the system is completely described by the $SO(3)$-valued extrinsic curvature $K^{i}_{\a}$ and its momentum $E_i^{\a}$, together with the field $\b$ and its momentum $\Pi$, for a total of 20 degrees of freedom. Then the physical degrees of freedom on the phase space are $20-2\times 7=6$,\footnote{The number of second-class constraints is even as it must be in order for them to be completely solved. Once the Dirac bracket is defined for the system, the second-class constraints can be considered as strong equations and the original variables are reintroduced in the theory by using a simple identity.} specifically four corresponding to the two polarizations of the graviton and two associated with the pseudoscalar field.

Before concluding the section, we wish to notice the self-consistency of the results achieved so far. One can verify that Eqs.~\Eq{sm} and \Eq{SH} do correspond to the supermomentum and super-Hamiltonian of the canonical theory based on the effective action Eq.~\Eq{lagr} for a minimally coupled, nonstandard scalar field, whose conjugate momentum is
\be
\Pi=\frac{3e}{2}\frac{1}{1+\b^2}n^\mu\p_\mu\b=\frac{3e}{2N}\frac{1}{1+\b^2}(\cL_t\b-N^\a\p_\a\b)\,.
\ee
Finally, we note that Eqs.~(\ref{s11}) and (\ref{t1}) are in agreement, after projection, with the solutions Eqs.~\Eq{pseu} and \Eq{tasol}.


\subsection{Nieh--Yan case}

The canonical analysis does not change much with respect to the previous section. The secondary constraints Eqs.~\Eq{s1}--\Eq{s4} become
\ba
\{C_{\a}^{(S)},H_D\}  &=& \frac{Ne}4\left(\p_\a\b-\frac{1}{6}S_\a\right)\,,\\
\{C_{\a}^{(T)},H_D\}  &=& \frac23NeT_\a\,,\\
\{C^{(T)},H_D\} &=& \frac23Ne(n\cdot T)\,,
\ea
leading to
\be
n\cdot T=0\,,\qquad T_\a=0\,,\qquad S_\a=6\p_\a\b\,.
\ee
The super-Hamiltonian turns out to be Eq.~\Eq{SH2}. One can show that the canonical theory stemming from Eq.~\Eq{lagr2} coincides with the one above.


\end{document}